\documentclass[12pt,preprint]{aastex}

\usepackage{amsmath}

\usepackage{graphics}
\usepackage{psfig}

\usepackage{color}

\newcommand{\bd}{\begin{displaymath}}
\newcommand{\ed}{\end{displaymath}}

\slugcomment{re-submitted to ApJ, on \today}

\begin{document}

\title{A general relativistic model of accretion disks with coronae surrounding Kerr black holes}

\author{Bei You\altaffilmark{1}, Xinwu Cao\altaffilmark{1}, and Ye-Fei Yuan\altaffilmark{2}}

\altaffiltext{1}{Key Laboratory for Research in Galaxies and
Cosmology, Shanghai Astronomical Observatory, Chinese Academy of
Sciences, 80 Nandan Road, Shanghai, 200030, China;
youbei@shao.ac.cn, cxw@shao.ac.cn} \altaffiltext{2}{Department of
Astronomy, University of Sciences and Technology of China, Hefei,
Anhui 230026, China; yfyuan@ustc.edu.cn}

\begin{abstract}
We calculate the structure of a standard accretion disk with corona
surrounding a massive Kerr black hole in general relativistic frame,
in which the corona is assumed to be heated by the reconnection of
the strongly buoyant magnetic fields generated in the cold accretion
disk. The emergent spectra of the accretion disk-corona systems are
calculated by using the relativistic ray-tracing method. We propose
a new method to calculate the emergent Comptonized spectra from the
coronae. The spectra of the disk-corona systems with a modified
$\alpha$-magnetic stress show that both the hard X-ray spectral
index and the hard X-ray bolometric correction factor $L_{\rm
bol}/L_{\rm X,2-10keV}$ increase with the dimensionless mass
accretion rate, which are qualitatively consistent with the
observations of active galactic nuclei (AGNs). The fraction
of the power dissipated in the corona decreases with increasing
black hole spin parameter $a$, which leads to lower electron
temperatures of the coronas for rapidly spinning black holes. The
X-ray emission from the coronas surrounding rapidly spinning black
holes becomes weak and soft. The ratio of the X-ray luminosity to
the optical/UV luminosity increases with the viewing angle, while
the spectral shape in the X-ray band is insensitive with the viewing
angle. We find that the spectral index in the infrared waveband
depends on the mass accretion rate and the black hole spin $a$,
which deviates from $f_\nu\propto\nu^{1/3}$ expected by the standard
thin disk model.
\end{abstract}



\section{Introduction}

Active galactic nuclei (AGNs) are powered by the accretion of gases
on to their central massive black holes, and the emission of the
accretion disks is responsible for the observed multi-band spectral
energy distributions (SEDs). It is suggested that the optically
thick and geometrically thin accretion disks are present in luminous
AGNs, and the observed UV/optical continuum emission of AGNs is the
thermal emission from the standard thin accretion disks
\citep*[e.g.,][]{1973A&A....24..337S,1978Natur.273..519S,1982ApJ...254...22M,1989ApJ...346...68S,1989MNRAS.238..897L,1990MNRAS.246..369L,2011MNRAS.415.2942C,2012arXiv1206.2569H}.
One feature in the spectra of such standard thin accretion disks
surrounding massive black holes is, $f_\nu\propto \nu^{1/3}$, in the
infrared waveband \citep{1999PASP..111....1K}. As the
accretion disk extends to the region very close to the black hole,
the general relativistic thin accretion disk models were developed
in the previous works
\citep*[e.g.,][]{1973blho.conf..343N,1974ApJ...191..499P,1989MNRAS.238..897L,2005ApJS..157..335L,2011arXiv1104.5499A},
which were applied successfully to fit the observed SEDs of AGNs and
the values of black hole spin parameter can be derived for some
sources \citep*[e.g.,][]{1990MNRAS.246..369L,2011MNRAS.415.2942C}.
The values of the spin parameter were estimated by the comparison of
the theoretical accretion disk spectra with the observed spectra in
some X-ray binaries
\citep{1997ApJ...482L.155Z,2009ApJ...701L..83S,2011ApJ...742...85G,2011MNRAS.414.1183K,2011CQGra..28k4009M}.
However, the temperature of the standard thin accretion disks is too
low to produce the observed power-law spectra of AGNs in the hard
X-ray waveband, and the accretion disk-corona model was therefore
suggested
\citep*[e.g.,][]{1979ApJ...229..318G,1991ApJ...380L..51H,1994ApJ...436..599S}.
In this scenario, the power-law hard X-ray spectra of AGNs are most
likely due to the inverse Compton scattering of soft photons on a
population of hot electrons in the corona above the disk. The soft
photons are emitted from the cold disk, and most of them pass
through the optically thin corona without being scattered, which
emerge as the observed optical/UV continuum emission of AGNs. A
small fraction of the photons from the cold disk are Compton
scattered by the hot electrons in the corona to the hard X-ray
energy band. The disk-corona model was extensively explored in many
previous works
\citep*[e.g.,][]{1991ApJ...380L..51H,1994ApJ...436..599S,2001ApJ...546..966K,2002ApJ...572L.173L,2009MNRAS.394..207C}.
Most gravitational energy is generated in the cold disk through the
turbulence, which is probably produced by the magnetorotational
instability \citep{1991ApJ...376..214B}. A fraction of the energy
generated in the cold disk should be transported to the corona. One
of the most promising mechanisms is the energy of the strongly
buoyant magnetic fields in the disk being transported vertically to
heat the corona above the disk with the reconnection of the fields
\citep*[e.g.][]{1998MNRAS.299L..15D,1999MNRAS.304..809D,2001MNRAS.328..958M}.
The temperature of the hot electrons in the corona is roughly around
10$^9$~K, which can successfully reproduce a power-law hard X-ray
spectrum as observed in AGNs
\citep*[e.g.][]{2003ApJ...587..571L,2009MNRAS.394..207C}.

The emergent spectrum of an accretion disk surrounding a rotating
black hole is influenced by the relativistic effects due to the
strong gravity field of the hole, which was extensively investigated
in many works
\citep[e.g.,][]{1985Ap&SS.113..181Z,1989MNRAS.238..897L,1991ApJ...376...90L,2005ApJS..157..335L,2009ApJ...699..513L}.
The iron $K\alpha$ fluorescence lines observed in some AGNs are
significantly asymmetric, characterized with a steep blue wing, an
extended red wing and the blueshifted emission line peak
\citep*[see][for a review, and the references
therein]{2007ARA&A..45..441M}.  These features can be modeled as
emission from the inner region of an accretion disk illuminated by
the external X-ray radiation with the Doppler boost effect and the
general relativistic effects (frame-dragging, gravitational redshift
and bending of the light) being properly considered
\citep*[e.g.,][]{1991ApJ...376...90L}. Such effects should also be
considered in the calculations of the emergent continuum spectrum of
an accretion disk surrounding a rotating black hole observed at
infinity
\citep*[e.g.,][]{1989MNRAS.238..897L,1997ApJ...482L.155Z,1991ApJ...376...90L,2005ApJS..157..335L,2009ApJ...699..513L}.
The ray-tracing method is widely adopted to derive the photon
trajectories bent by the strong gravity field of the black hole in
the calculations of the emergent spectra emitted from the disk
\citep*[e.g.,][]{1983mtbh.book.....C,1994ApJ...421...46R,1998tx19.confE.391C,2004A&A...424..733F,2009ApJ...699..722Y,2011arXiv1104.5499A}.
The values of black hole spin parameter can be estimated by the
comparison of the observed spectra with the theoretical model
calculations \citep*[e.g.,][]{2010ApJ...723..508Y}. Besides the
radiation from the disk, the emergent spectrum of the corona is also
affected by these general relativistic effects, which is more
complicated than the disk case due to the complexity of the
radiation transfer of Comptonized photons in the geometrically thick
corona \citep[e.g.,][]{2010ApJ...712..908S}.

In this work, we calculate the structure of a thin accretion disk
with corona surrounding a massive Kerr black hole in general
relativistic frame, in which the corona is assumed to be heated by
the reconnection of the magnetic fields generated by buoyancy
instability in the cold accretion disk.  The emergent spectra of the
accretion disk-corona systems are calculated by using the
relativistic ray-tracing method. The emergent Comptonized spectra
from the coronae are calculated by dividing layers in the coronae.
We summarize the calculations of the disk-corona structure and its
emergent spectrum in Sections 2-4. The numerical results of the
model calculations and the discussion are given in Sections 5 and 6.

\section{The accretion disk-corona structure}

In the Boyer-Lindquist coordinates $(t, r, \theta, \phi)$ with
natural units $G=c=1$, the Kerr metric is
\begin{equation}\label{metric}
    ds^{2}=-e^{2\nu}dt^2+e^{2\psi}(d\phi-\omega dt)^2+e^{2\mu_1}dr^2+e^{2\mu_2}d\theta^2
\end{equation}
where
\begin{displaymath}\label{e}
e^{2\nu}=\frac{\Sigma \Delta}{\xi}, \,\,\,
e^{2\psi}=\frac{\xi \sin ^2 \theta}{\Sigma}, \,\,\,
e^{2\mu _1}=\frac{\Sigma}{\Delta}, \,\,\,
e^{2\mu_2}=\Sigma, \,\,\,
\omega=\frac{2M_{\rm bh}ar}{\xi},  \,\,\,
\end{displaymath}
\begin{equation}\label{delta}
\Delta=r^2-2M_{\rm bh}r+a^2,  \,\,\, \Sigma=r^2+a^2\cos ^2\theta, \,\,\,
\xi=(r^2+a^2)^2-a^2\Delta\sin ^2\theta,
\end{equation}
$M_{\rm bh}$ is the mass of black hole, and $a$ is the black hole
spin parameter $(-1\leq a \leq 1)$. The event horizon of the black
hole is
\begin{equation}\label{horizon}
    r_+=M_{\rm bh}+(M_{\rm bh}^2-a^2)^{1/2}.
\end{equation}
The Keplerian velocity of the matter moving around the black hole in
circular orbits in the equatorial plane at radius $R$ is given by
\begin{equation}\label{keplerian}
    \Omega_{\rm K} =\pm {\frac {M_{\rm bh}^{1/2}}{R^{3/2} \pm aM_{\rm bh}^{1/2}}},
\end{equation}
where the upper sign refers to direct orbits, i.e., corotating with
the black hole spinning direction, and the lower sign refers to
retrograde orbits, i.e., counterrotating with the black hole
spinning direction. The innermost of a geometrically thin accretion
disk is assumed to extend to the marginally stable circular orbit of
the black hole, which is given by
\begin{equation}
  r_{\rm ms}=M_{\rm bh}\{3+Z_2\mp [(3-Z_1)(3+Z_1+2Z_2)]^{1/2}\},
\end{equation}
where
\begin{displaymath}
Z_1=1+\left(1-{\frac {a^2}{M_{\rm bh}^2}}\right)^{1/3}\left[\left(1+{\frac
{a}{M_{\rm bh}}}\right)^{1/3}+\left(1-{\frac {a}{M_{\rm bh}}}\right)^{1/3}\right],
\end{displaymath}
and
\begin{displaymath}
Z_2=\left({\frac {3a^2}{M_{\rm bh}^2}}+Z_1^2\right)^{1/2}.
\end{displaymath}

In this work, we consider the relativistic accretion disk with
corona, in which the cold disk is optically thick and geometrically
thin. The circular motion of the matter in the accretion disk is
assumed to be Keplerian, as that in the standard thin disk model
\citep{1973A&A....24..337S}. The formulation of the relativistic
accretion disk model is similar to the standard thin disk model, in
which some general relativistic correction factors $A$, $B$, $C$,
$D$, and $L$, are included \citep{1973blho.conf..343N,1974ApJ...191..499P}. The
values of all these quantities approach unity in the region far from
the black hole \citep*[see][for a review]{2011arXiv1104.5499A}.

The gravitational power of the matter dissipated in unit surface
area of the accretion disk is
\begin{equation}
Q_{\rm dissi}^{+}={\frac {3GM_{\rm bh}\dot{M}}{8\pi R^{3}}}{\frac {L}{BC^{1/2}}},
\label{q_disk_plus}
\end{equation}
where $\dot{M}$ is the mass accretion rate of the disk. A fraction
of the dissipated power is transported to the corona most probably
by the magnetic fields generated in the disk. The fields are
strongly buoyant, and the corona above the disk is heated with the
reconnection of the magnetic fields \citep{1998MNRAS.299L..15D}. The
soft photons emitted from the disk are Compton scattered by the hot
electrons in the corona to high energy, which is the main cooling
process in the corona. About half of the scattered photons are
intercepted by the disk, part of which are reflected and the
remainder are re-radiated as blackbody radiation
\citep{1999MNRAS.303L..11Z}. Thus, the energy equation for the cold
disk is
\begin{equation}
Q_{\rm dissi}^{+}-Q_{\rm cor}^{+}+\frac{1}{2}(1-f)Q_{\rm
cor}^{+}=\frac{4\sigma T_{\rm d}^{4}}{3\tau} \label{energy_disk}
\end{equation}
where $T_{\rm d}$ is the interior temperature in the mid-plane
of the disk, and $\tau=\tau_{\rm es}+\tau_{\rm ff}$ is the optical
depth in the vertical direction of the disk, and the reflection
albedo $f=0.15$ is adopted in all our calculations. The power
transported from the disk to the corona is
\citep{1998MNRAS.299L..15D}
\begin{equation}\label{corona_plus}
    Q_{\rm cor}^{+}=p_{\rm m}\upsilon_{\rm p}=\frac{B^{2}}{8\pi}\upsilon_{\rm p}
\end{equation}
where $p_{\rm m}$ is the magnetic pressure in the disk, and
$\upsilon_{\rm p}$ is the vertically rising speed of the magnetic
fields in the accretion disk.  The rising speed $\upsilon_{\rm p}$
is assumed to be proportional to internal Alfven velocity, i.e.,
$\upsilon_{\rm p}=b\upsilon_{\rm A}$, in which $b$ is of the order
of unity for extremely evacuated magnetic loops. The strength of the
magnetic fields in the disk is crucial in our present investigation,
which determines the ratio of power dissipated in the corona
to that in the disk. This ratio can be constrained by the observed
ratio of the X-ray to bolometric luminosities, which seems to
support that the values of $b$ should not deviate significantly from
unity in most AGNs, though a low value of $b$ (e.g. $b\la 0.1$)
cannot be ruled out in few individual AGNs with extremely low
$L_{\rm X}/L_{\rm bol}$
\citep*[see][]{2007MNRAS.381.1235V,2009MNRAS.394..207C}. The
detailed physics for generating magnetic fields in the accretion
disks is still quite unclear, and we assume the magnetic stress
tensor to be related to the pressure of the disk, as done in many
previous works
\citep*[e.g.,][]{1981ApJ...247...19S,1984ApJ...277..312S,1984ApJ...287..761T,2009ApJ...704..781H}.
\citet{2009MNRAS.394..207C} constructed accretion disk-corona models
with different magnetic stress tensors, and found that the model
calculations can qualitatively reproduce the observational features
that both the hard X-ray spectral index and the hard X-ray
bolometric correction factor $L_{\rm bol}/L_{\rm X,2-10 keV}$
increase with the Eddington ratio, if $\tau_{r\varphi}=p_{\rm
m}=\alpha \sqrt{p_{\rm gas}p_{\rm tot}}$ ($p_{\rm tot}=p_{\rm
gas}+p_{\rm rad}$ is adopted). We use this magnetic stress tensor in
most of our calculations in this work.



The angular momentum equation for the gas in the disk is
\citep{1973blho.conf..343N}
\begin{equation}
4\pi H_{\rm d}\tau_{r\varphi}=\dot{M}\sqrt{\frac{GM_{\rm bh}}{
R^{3}}}\frac{F}{D}, \label{angular}
\end{equation}
where $H_{\rm d}$ is the scale height of the accretion disk and the
factor $F$ is first introduced in \cite{1995ApJ...450..508R} in the
form of a function $D$. In the vertical direction of the accretion
disk, the hydrostatic equilibrium is assumed. The vertical pressure
gradient being balanced with the vertical component of the gravity
leads to \citep{1995ApJ...450..508R,2006ApJ...652..518M}
\begin{equation}\label{balance}
 \frac{dp}{dz}=-\frac{\rho GM_{\rm bh}z}{R^{3}}
  \frac{J}{C}
\end{equation}
which is valid for the geometrically thin disk, and $J=1-\frac{4a}{r^{3/2}}+\frac{3a^{2}}{r^{2}}$. Thus the scale
height of the disk is given by
\begin{equation}\label{height}
    H_{\rm d}=c_{\rm s}\left(\frac{R^{3}}{GM_{\rm
    bh}}\frac{C}{J}\right)^{1/2},
\end{equation}
where the sound speed $c_{\rm s}=[(p_{\rm tot}+p_{\rm
m})/\rho]^{1/2}$.


The continuity equation of the disk is
\begin{equation}
\dot{M}=-4\pi RH_{\rm d}\rho V_{r}D^{1/2}, \label{continuity}
\end{equation}
where $\rho(R)$ is the mean density of the disk, $V_{r}(R)$ is the
radial velocity of the accretion flow at radius $R$, and the mass
accreted in the corona is neglected.

The state equation of the gas in the disk is
\begin{equation}\label{state}
p_{\rm tot}=p_{\rm gas}+p_{\rm rad}=\frac{\rho kT_{\rm d}}{\mu
m_{\rm p}}+\frac{1}{3}a_0T_{\rm d}^{4}
\end{equation}
where $a_0$ is the radiation constant, the mean atomic mass $\mu =
(1/\mu_{\rm i}+1/\mu_{\rm e})^{-1}$. We assume the plasma to consist
of 3/4 hydrogen and 1/4 helium, i.e., $\mu_{\rm i}=1.23$ and
$\mu_{\rm e}=1.14$.

The ions and electrons in the corona are heated by the reconnection
of the magnetic fields rising from the disk, of which the cooling is
dominated by the inverse Compton scattering of the soft photons from
the disk, the synchrotron radiation, and the bremsstrahlung
radiation. The energy equation of the two-temperature corona is
given by  \citep{1998MNRAS.299L..15D}
\begin{equation}\label{energy_cor}
   Q_{\rm cor}^{+}=Q_{\rm cor}^{\rm ie}+\delta Q_{\rm cor}^{+}=F_{\rm
   cor}^{-},
\end{equation}
where $F_{\rm cor}^{-}=F_{\rm syn}^{-}+F_{\rm brem}^{-}+F_{\rm
comp}^{-}$ is the cooling rate in unit surface area of the corona,
and $Q_{\rm cor}^{\rm ie}$ is the energy transfer rate from the ions
to the electrons in the corona via Coulomb collisions, which is
given by \citet{1983MNRAS.204.1269S}. We adopt the fraction of the
energy directly heating the electrons $\delta =0.5$ in all our
calculations \citep{1997ApJ...486L..43B,2000ApJ...529..978B}. The
cooling terms $F_{\rm syn}^{-}$ and $F_{\rm brem}^{-}$ are the
functions of the magnetic field strength, the number density and
temperature of the electrons in the corona, which are taken from
\citet{1995ApJ...452..710N}. The cooling rate $F_{\rm Comp}^{-}$ of
the Compton scattering in the corona is calculated by using the
approach suggested by \citet{1990MNRAS.245..453C}.


The state equation of the gas in the optically thin corona is
\begin{equation}\label{pressure_cor}
    p_{\rm cor}=\frac{\rho_{\rm cor}kT_{\rm i}}{\mu_{\rm i}m_{\rm p}}+\frac{\rho_{\rm cor}kT_{\rm e}}{\mu_{\rm e}m_{\rm p}}+p_{\rm
    cor,m},
\end{equation}
where $T_{\rm i}$ and $T_{\rm e}$ are the temperatures of the ions
and electrons in the two-temperature corona, and the magnetic
pressure $p_{\rm cor,m}=B_{\rm cor}^{2}/8\pi$. Given uncertainty of
the magnetic fields in the corona, we simply assume the magnetic
pressure to be equipartition with the gas pressure in the corona
\citep{2009MNRAS.394..207C}. The cooling of the corona is always
dominated by the inverse Compton scattering, and the strength of the
fields in the corona mainly affects the synchrotron radiation. Our
final results are almost independent of the value of the field
strength adopted, except for the spectra in radio wavebands. The
scale height of the corona $H_{\rm cor}$ can be estimated based on
the assumption of static hydrodynamical equilibrium in the vertical
direction.

As done in the work of \citet{2009MNRAS.394..207C}, we need to
specify the ion temperature of the corona in the calculations of the
disk-corona structure. It is found that the ion temperature $T_{\rm
i}$ has little influence on the X-ray spectra, and
\cite{2003ApJ...587..571L} pointed out that the temperature of the
ions in the corona is always in the range of $\sim$ 0.2-0.3 virial
temperature $\left(T_{\rm vir}=GM_{\rm bh}m_{\rm p}/kR\right)$.
In this work, the definition of the virial temperature is slightly
different,
\begin{equation}\label{virial}
   \frac{3}{2}kT_{\rm vir}=\frac{1}{2}m_{\rm p}\Omega_{\rm
   k}^{2}R^{2},
\end{equation}
where $\Omega_{\rm K}$ is the Keplerian velocity of the gas in the
disk at radius $R$.
In this work, we adopt $T_{\rm i}=0.9T_{\rm vir}$ in all our
calculations, which is almost equivalent to the value suggested by
\citet{2003ApJ...587..571L}.


It is still quite unclear on the outer radius of the accretion disk.
One conventional estimate of the outer radius of the disk is based
on the assumption that the disk is truncated where it becomes
gravitational unstable \citep*[e.g.,][]{2008A&A...477..419C}. The
outer radius of the disk is estimated with the Toomre
parameter\citep{1964ApJ...139.1217T,1965MNRAS.130...97G},
\begin{equation}\label{rmax}
        Q_{\rm Toomre}={\frac{\Omega_{\rm K}c_{\rm s}}{2\pi G\rho H_{\rm d}}}=1.
\end{equation}

Given the values of the black hole mass $M_{\rm BH}$, the spin
parameter $a$, the viscosity parameter $\alpha$, and the mass
accretion rate $\dot{M}$, the disk structure can be derived by
solving Equations (\ref{q_disk_plus})-(\ref{corona_plus})
numerically. The structure of the corona is then available based on
the derived disk structure by solving Equations
(\ref{energy_cor})-(\ref{virial}).


\section{Ray tracing method}

The trajectory of the photons emitted from the inner region of the
accretion disk-corona system is bent by the strong gravity of the
black hole. The observed emergent spectrum of such an accretion
disk-corona system surrounding a Kerr black hole can be calculated
with the ray tracing method. We summarize the method briefly in this
section.


In the Kerr spacetime, there are four constants for the motion of a
photon \citep{1968PhRv..174.1559C,2011arXiv1104.5499A}, three of which can determine the
trajectory of the photon, namely the energy at infinity,
\begin{equation}\label{infinity}
    E=-p_{\rm t},
\end{equation}
the effective angular momentum,
\begin{equation}\label{momentum}
    L_{z}=p_\phi ,
\end{equation}
and the Carter constant,
\begin{equation}\label{carter}
    Q=p_\theta ^2-a^2E^2\cos^2\theta +L_{z}^2\cot^2\theta ,
\end{equation}
where $p_\mu$ is the four-momentum of the photon. Combining
Equations (\ref{infinity})-(\ref{carter}) and the expressions of a
photon's four-momentum in \cite{1968PhRv..174.1559C}, we derive the
following equations describing the trajectory of a photon in the
Kerr spacetime:
\begin{eqnarray}
\Sigma\frac{dr}{d\sigma} &=& \pm\sqrt{R(r)}, \\
\Sigma\frac{d\theta}{d\sigma} &=& \pm\sqrt{\Theta(\theta)}, \\
\Sigma\frac{d\phi}{d\sigma} &=& \frac{L_z}{\sin^2\theta}+\frac{2arE-L_za^2}{\Delta}, \\
\Sigma\frac{dt}{d\sigma} &=&
-a(aE\sin^2\theta-L_z)+\frac{(r^2+a^2)[E(r^2+a^2)-L_za]}{\Delta},
\end{eqnarray}
where \begin{eqnarray}
R(r) &=& [E(r^2+a^2)-L_za]^2-\Delta [(L_z-aE)^2+Q], \\
\Theta(\theta) &=&
Q-(-a^2E^2+\frac{L_z^2}{\sin^2\theta})\cos^2\theta.
      \end{eqnarray} \\
Here $\sigma$ is the affine parameter along the trajectory. The
signs in the above equations must be the same as the signs of $dr$,
$d\theta$ respectively, and they change at the turning point in $r$
or $\theta$, i.e., $R(r)=0$ or $\Theta(\theta)=0$. As the trajectory
of a photon is independent of its energy $E$,
\cite{1973ApJ...183..237C} introduce two dimensionless parameters
being conserved along the trajectory,
\begin{equation}\label{dimensionless_constant}
\lambda\equiv \frac{L_z}{E},  \,\,\,
q\equiv\frac{Q}{E^2}
\end{equation}
and define another affine parameter $dP=E/\Sigma d\sigma$. The
equations describing the trajectory of the photon can be re-written
as
\begin{eqnarray}
      dr &=& \pm\sqrt{\widetilde{R}(r)}dP, \\
      d\theta &=& \pm\sqrt{\widetilde{\Theta}(\theta)}dP, \\
      d\phi &=& \left[\frac{\lambda}{\sin^2\theta}+\frac{2ar-\lambda a^2}{\Delta} \right]dP, \\
      dt &=& \left[\frac{(r^2+a^2)^2}{\Delta}-a^2\sin^2\theta -\frac{2ar\lambda}{\Delta}
      \right]dP,
    \end{eqnarray}
where $\widetilde{R}(r)\equiv R(r)/E^2$,
$\widetilde{\Theta}(\theta)\equiv \Theta(\theta)/{E^2}$. Integrating
the null geodesic equations of the photon along the path from the
initial location $(r_{\rm em},\theta_{\rm em},\phi_{\rm em})$ to the
observer at infinity with position $(r_{\rm obs}, \theta_{\rm
obs},\phi_{\rm obs})$, the trajectory can be determined by the
following integral equations
\begin{equation}\label{geodesic}
    \pm\int\limits_{r_{\rm em}}^{r_{\rm obs}}\frac{dr}{\sqrt{\widetilde{R}(r)}}=\pm\int\limits_{\theta_{\rm em}}^{\theta_{\rm obs}}\frac{d\theta}{\sqrt{\widetilde{\Theta}(\theta)}}=P,
\end{equation}
and
\begin{equation}\label{phi}
    \phi_{\rm obs}-\phi_{\rm em}=\pm\int\limits_{P_{\rm em}}^{P_{\rm obs}}\left(\frac{\lambda}{\sin^2\theta}+\frac{2ar-a^2\lambda}{\Delta}\right)dP,
\end{equation}
where $P_{\rm em}$, $P_{\rm obs}$ are functions of $\theta$ or $r$,
corresponding to the upper limit value of the $\theta$-integral
taken as $\theta_{\rm em}$, $\theta_{\rm obs}$ respectively, and the signs in the above integrals must be the same as those of $dr$ and $d\theta$ to guarantee the integral $P$ is always positive and increasing along a photon's trajectory. In the
calculations, we set $r_{\rm obs}=\infty$, and $\phi_{\rm obs}=0$.
The detailed explanation on the integrals can be found in some
previous works
\citep{1983mtbh.book.....C,1998tx19.confE.391C,2005ApJS..157..335L}.

The image of the accretion disk-corona observed at an angle
$\theta_{\rm obs}$ with respect to the axis of the black
hole at infinity can be described by the two impact parameters,
\begin{equation}\label{alpha}
    \alpha=-\left(\frac{rp^{(\phi)}}{p^{(t)}}\right)_{r\rightarrow \infty }=-\frac{\lambda}{\sin \theta
    _{\rm obs}},
\end{equation}
and
\begin{equation}\label{beta}
    \beta=\left(\frac{rp^{(\theta)}}{p^{(t)}}\right)_{r\rightarrow \infty }=(q+a^2\cos^2\theta_{\rm obs}-\lambda
    ^2\cot^2\theta_{\rm obs})^{1/2}=p_{\rm obs},
\end{equation}
where $\alpha$ is the apparent displacement of the image
perpendicular to the projected axis of the black hole, and $\beta$
is the apparent parallel displacement. The vector $p^{(a)}$ is the
four-momentum in the local nonrotating frame(LNRF). The observed
image of the accretion disk-corona can be calculated with the ray
tracing method described above, provided that the local spectrum of
the disk-corona is known. For a given point ($\alpha$, $\beta$) in
the image seen by the observer at ($\infty, \theta_{\rm obs}, 0$),
the two constants of motion $\lambda$ and $q$ can be derived with
Equations (\ref{alpha}) and (\ref{beta}). Then the trajectory
integral $P$, the radius $r_{\rm em}$, and the azimuthal angle
$\phi_{\rm em}$, can be calculated by using Equations
(\ref{geodesic}) and (\ref{phi}).



\section{Radiation transfer and emergent spectrum}

In this work, the photons cannot move through the equatorial plane
because of the optically thick accretion disk, and therefore we
neglect such trajectories. For the photons passing through the
corona, the optical depth for the scattering of photons in the
corona in the curved space can be calculated with the ray tracing
method.
Along the trajectory of the photon in the corona, the differential
optical depth is
\begin{equation}\label{depth}
    d\tau=\sigma_Tn_{\rm e}dl,
\end{equation}
where $\sigma_{\rm T}$ is the Thompson cross-section, $n_{\rm e}$ is the
electrons number density in the corona at $(R,~z)$, and
\begin{equation}\label{dl}
    dl=e^\nu \Sigma dP=\pm \frac{e^\nu
    \Sigma}{\sqrt{\tilde{\Theta}(\theta)}}d\theta.
\end{equation}
Integrating along the photon trajectory of the photon in the corona,
we have
\begin{equation}\label{tau}
    \tau=\pm\int\limits_{\theta_{\rm em}}^{\theta_{\rm obs}}\frac{\sigma_{\rm T}n_{\rm e}e^\nu
    \Sigma}{\sqrt{\tilde{\Theta}(\theta)}}d\theta,
\end{equation}
where $d\theta=\pm\sqrt{\tilde{\Theta}(\theta)}dP$.


As the accretion disk is geometrically thin, we approximate
$\theta_{\rm em}=\pi/2$ in calculating the emergent spectrum of the
disk. The radiative flux from the accretion disk observed at
infinity is given by
\begin{equation}\label{flux_disk}
    F(\nu_{\rm obs})=\int e^{-\tau}B_{\nu_{\rm obs}}d\Omega=\int e^{-\tau}B_{\nu_{\rm
    obs}} \frac{d\alpha d\beta}{D^2},
\end{equation}
where $B_{\nu_{\rm obs}}$ is the intensity at observed frequency
$\nu_{\rm obs}$, $d\Omega=d\alpha d\beta/D^2$ is the solid angle
subtended by the image of the disk.

The gravitational redshift effect and Doppler redshift can be included by using the Lorentz invariant $I/\nu^3$ \citep{1979rpa..book.....R}, so the
emergent spectrum can be calculated with
\begin{equation}\label{redshift}
    F({\nu_{\rm obs}})=\int g^3e^{-\tau}B_{\nu_{\rm e}}\frac{d\alpha d\beta}{D^2}
\end{equation}
where $B_{\nu_e}$ is the the intensity of the photons from the disk
at emitting frequency $\nu_e$. Integrating over the disk image, the
emergent spectrum of the disk is available with
\begin{equation}\label{Ldisk}
    L_{\rm disk}(\nu_{\rm obs})=4\pi D^2F({\nu_{\rm obs}})=4\pi \int g^3e^{-\tau}B_{\nu_{\rm e}}d\alpha
    d\beta.
\end{equation}
The redshift factor is defined by \citep{1973ApJ...183..237C}
\begin{equation}\label{g}
    g\equiv\frac{\nu_{\rm obs}}{\nu_{\rm e}}=\frac{p_\mu u^\mu|_{\rm obs}}{p_\mu
    u^\mu|_{\rm em}},
\end{equation}
where $p_\mu$ is the four-momentum of the photon
\citep{1968PhRv..174.1559C},
\begin{equation}\label{fourmomentum}
    p_\mu=(p_t,p_r,p_\theta,p_\phi)=(-1,\pm\frac{\sqrt{\widetilde{R}(r)}}{\Delta},\pm\sqrt{\widetilde{\Theta}(\theta)},\lambda)E.
\end{equation}
The four-velocity of the fluid in the locally non-rotating frame
(LNRF) $u^\mu_{\rm obs}$ and the local rest frame (LRF) $u^\mu_{\rm
em}$ are given by
\begin{equation}\label{fluidobs}
      u_{\rm obs}^\mu=(1,0,0,0),
\end{equation}
and
\begin{equation}\label{fluidem}
      u_{\rm em}^\mu=(\gamma_r\gamma_\phi e^{-\nu},\gamma_r\beta_re^{-\mu_1},0,\gamma_r\gamma_\phi (\omega e^{-\nu}+\beta_\phi
      e^{-\psi})),
\end{equation}
respectively \citep{2009ApJ...699..722Y}.

The corona is assumed to rotate with the disk at Keplerian
velocity. The calculations of the emergent spectrum of the corona is
slightly different from that of the accretion disk, because the
redshift factor $g$ and the trajectory of a photon depends on the
location of the photon emitted $(r_{\rm em},~\phi_{\rm em},~z_{\rm
em})$ in the corona. To calculate the redshift factor $g$ and the
trajectory of such a photon, we numerically solve Equations
(\ref{geodesic}) and (\ref{phi}) to derive the impact parameters
$\alpha$ and $\beta$.
%
We divide the corona into small volumes with $dh d\epsilon$ where $d\epsilon$ is element area and $dh$ is the thickness of a layer, and the emergent spectrum of each volume is available by calculating the
radiation transfer with the ray tracing method, if the emissivity of
the corona is available. For simplicity, we assume the radiation of
the corona to be isotropic and the emissivity is homogeneous in the
vertical direction of the corona.


The synchrotron and bremsstrahlung emissivities are taken from
\citet{1995ApJ...452..710N}.
The spectrum of synchrotron and bremsstrahlung emission from the
corona is available by summing up the contribution from these small
volumes,
\begin{equation}
L_{\rm syn, brem}=4\pi\int
g_{h}^{3}e^{-\tau_{h}}(\varepsilon_{\nu_{\rm e}}^{\rm
syn}+\varepsilon_{\nu{\rm e}}^{\rm brem})d\alpha d\beta,
\label{synbrem}
\end{equation}
where the effects of the absorption and the scattering in the corona
are considered as described at the beginning of this section.

In the calculations of the Comptonization in the corona, we assume
the corona to be a parallel plane \citep*[see][for the
details]{2009MNRAS.394..207C}. The mean probability of the soft
photons injected from the cold disk experiencing first-order
scattering in the corona is
\begin{equation}
P_{1}=2\int\limits_{0}^{1}(1-e^{-\tau_0/\cos\theta})\cos\theta{
d}\cos\theta,
 \label{prob_scat1}
\end{equation}
where the constant specific intensity of the soft photons from the
disk is assumed, $\theta$ is the angle of the motion of the soft
photons with respect to the vertical direction of the disk, and
$\tau_0=\sigma_{T}n_{\rm e}H_{\rm cor}$ is the optical depth of the
corona for electron scattering in the vertical direction. We can
calculate the first-order Comptonized spectra of the corona
$F_{\nu,1}^{\rm Comp}$ using the method suggested by
\citet{1990MNRAS.245..453C} with the probability given by Equation
(\ref{prob_scat1}), if the density and the temperature of electrons
in the corona, and the incident spectrum of the cold disk are known.
As we have assumed the scattered photons to be isotropic, the
emissivity of the first-order Compton scattered photons in the layer
with the height between $h$ and $h+dh$ can be approximated as
$\epsilon_{\nu,1}^{\rm Comp}(R_{\rm d})=F_{\nu,1}^{\rm Comp}dh/H_{\rm cor}$. The first-order Compton scattered photons will pass
though the corona before arriving the observer. Using the above
described ray tracing approach, we can calculate the emergent
first-order Comptonized spectrum of the corona.

The mean probability for these first-order scattered photons
experiencing the second-order scattering can be estimated as
\begin{equation}
P_{2}={1\over2}\int\limits_{0}^{1}{\rm
d}\xi\int\limits_{0}^{1}[1-e^{-\xi\tau_0/\cos\theta}+1-e^{-(1-\xi)\tau_0/\cos\theta}]{\rm
d}\cos\theta, \label{prob_scat2}
\end{equation}
where $\xi=z/H_{\rm cor}$, and the first-order scattered photons are
assumed to be radiated isotropically. The two terms in
the integration are for downward-moving and upward-moving photons emitted at height $z$,
respectively. The probability of the
scattered photons experiencing next higher order scattering
approximates to $P_2$, so we simply adopt $P_{n}=P_2$ for $n>2$.
Thus, the $n$th-order Comptonized spectra $F_{\nu,n}^{\rm Comp}$ and
emissivity $\epsilon_{\nu,n}^{\rm Comp}$ can be derived with the
same method for calculating the first-order Comptonization, in which
the $(n-1)$th-order Comptonized spectra are used instead of the
incident spectrum from the cold disk used for the calculation of the
first-order Comptonization. In the same way, the emergent spectra of
high order Comptonized spectra are derived, and the total
Comptonized spectrum of the corona is available by summing up the
contribution from the whole corona,
\begin{equation}\label{comp}
L_{\rm Comp}=4\pi\int g_{h}^{3}(\varepsilon_{\nu,1}^{\rm
Comp}+\varepsilon_{\nu,2}^{\rm Comp}+...+\varepsilon_{\nu,n}^{\rm
Comp})d\alpha d\beta.
\end{equation}


\section{Results}

As described in Section 2, we can calculate the structure of an
accretion disk with corona surrounding a rotating Kerr black hole,
if the black hole mass $M_{\rm bh}$, the spin parameter $a$, and the
dimensionless accretion rate $\dot{m}$ are specified. Based on the
derived structure of the accretion disk-corona system, the emergent
spectrum observed at infinity with an inclination angle $\theta_{\rm
obs}$ is calculated with the ray tracing method described in
Sections 3 and 4, in which all the general relativistic effects are
considered. The black hole mass $M_{\rm bh}=10^{8}M_{\odot}$ is
adopted in most of our calculations for AGNs. The
dimensionless mass accretion rate is defined as
$\dot{m}=\dot{M}/\dot{M}_{\rm Edd}$ ($\dot{M}_{\rm Edd}=L_{\rm
Edd}/0.1c^{2}$) independent of black hole spin.

\begin{figure}
\epsscale{0.9} \plotone{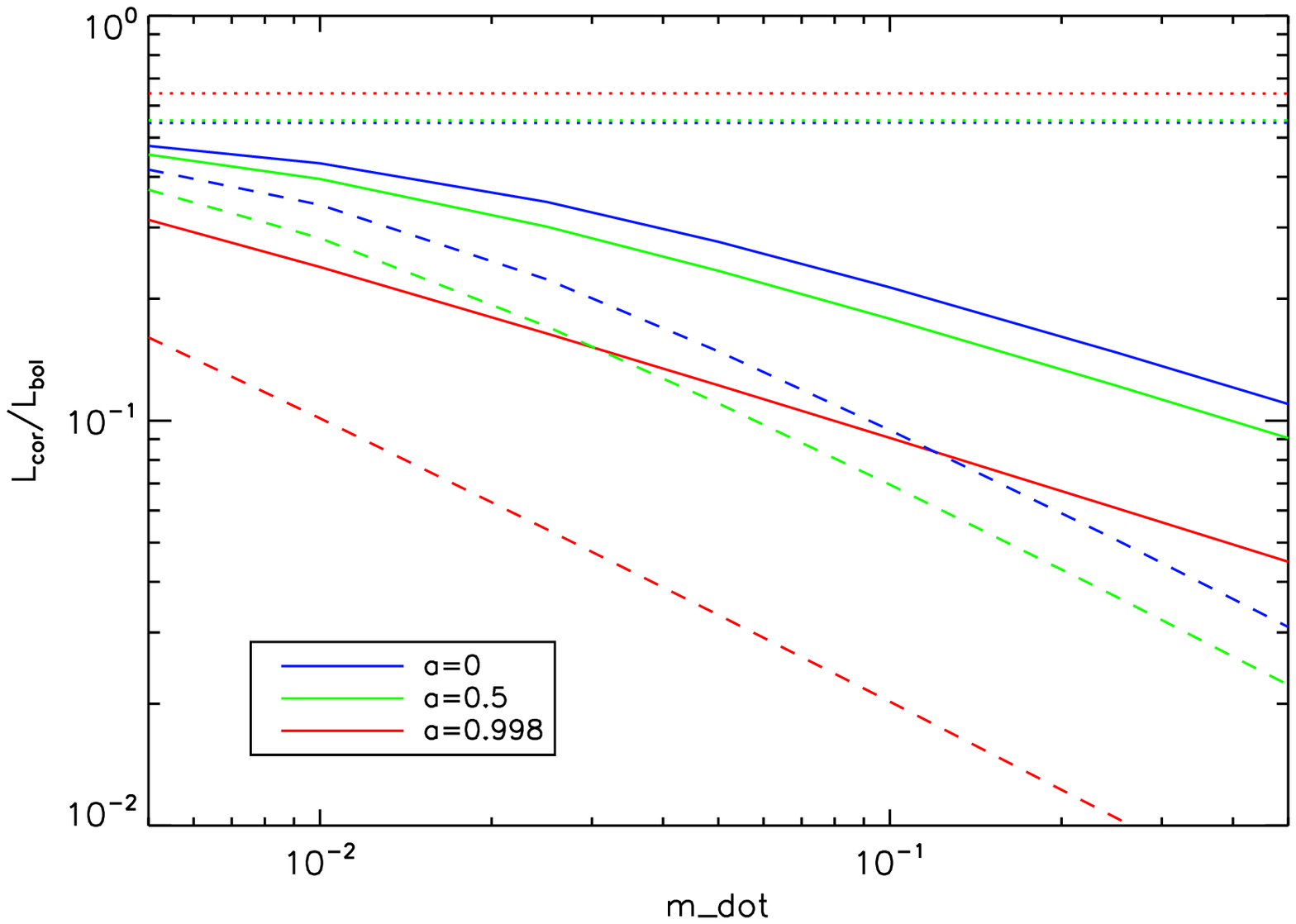} \caption{The ratios $L_{\rm
cor}/L_{\rm bol}$ as functions of accretion rate $\dot{m}$ with
different magnetic stress tensors. The color lines represent
different values of black hole spin parameter, i.e., $a=0$ (blue),
$0.5$ (green), and $0.998$ (red).  The different type lines
correspond to different magnetic stress tensors, i.e.,
$\tau_{r\varphi}=\alpha p_{\rm tot}$ (dotted lines),
$\tau_{r\varphi}=\alpha \sqrt{p_{\rm gas}p_{\rm tot}}$ (solid
lines), and $\tau_{r\varphi}=\alpha p_{\rm gas}$ (dashed
lines).}\label{fig1}
\end{figure}

We plot the ratios of the power dissipated in the corona to the
total released gravitational power $L_{\rm cor}/L_{\rm bol}$ as
functions of the accretion rate $\dot{m}$ with different values of
the black hole spin parameter $a$ in Figure \ref{fig1}. It is found
that the dependence of the ratio $L_{\rm cor}/L_{\rm bol}$ with
$\dot{m}$ is quite different for the cases with different magnetic
stress tensors, which are qualitatively consistent with those
obtained in \citet{2009MNRAS.394..207C} for the Newtonian accretion
disk corona model. We find that the ratio $L_{\rm cor}/L_{\rm bol}$
decreases with increasing spin parameter $a$ with either
$\tau_{\rm r\varphi}=\alpha P_{\rm gas}$ or $\tau_{\rm
r\varphi}=\alpha\sqrt{P_{\rm gas}P_{\rm tot}}$ if all other
parameters are fixed, which means that the efficiency of the energy
transportation from the disk to the corona is low for a rapidly
spinning black hole independent of the detailed magnetic stress
tenor adopted. However the model with $\tau_{\rm
r\varphi}=\alpha P_{\rm tot}$ appears to show the opposite trend
that $L_{\rm cor}/L_{\rm bol}$ reaches large values for a rapidly
spinning black hole.

\begin{figure}
\epsscale{0.9} \plotone{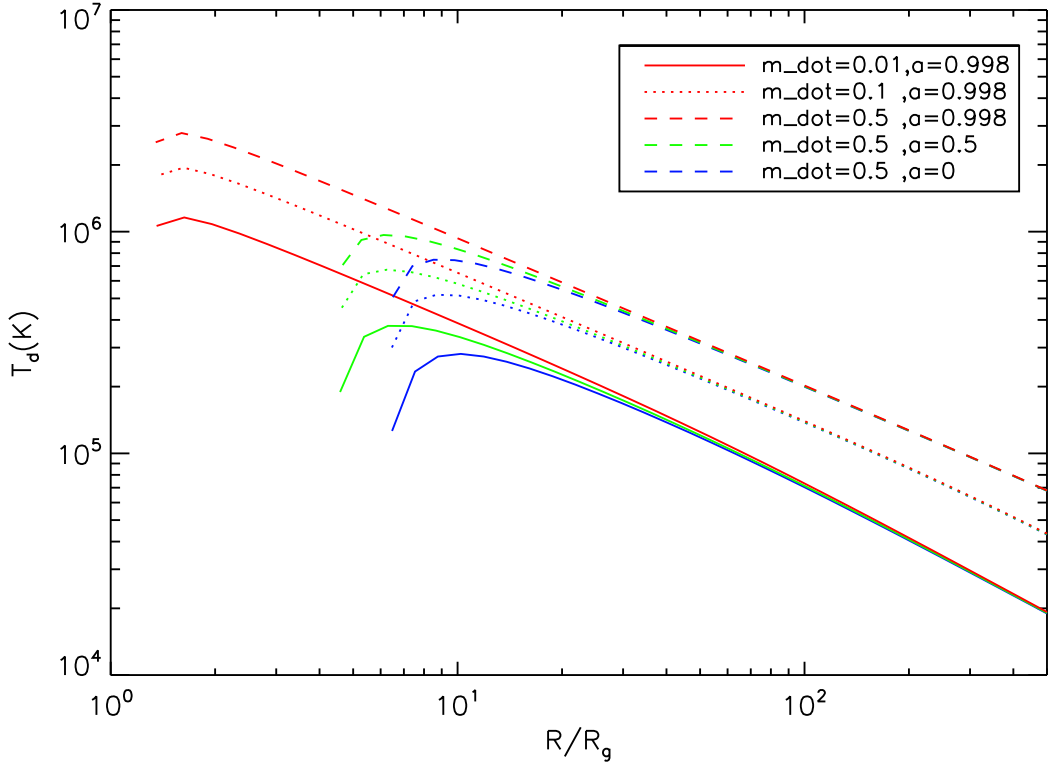} \caption{The temperatures
in the mid-plane of the accretion disks as functions of radius $R$
in units of $R_{\rm g}=GM_{\rm bh}/c^{2}$ with different values of
accretion rate $\dot{m}$ and black hole spin parameter $a$. The
colored lines represent the model results with different values of
black hole spin parameter, i.e., $a=0$ (blue), $0.5$ (green), and
$0.998$ (red).  The solid lines represent the results calculated
with $\dot{m}=0.01$, while the dotted and dash lines are for
$\dot{m}=0.1$ and $0.5$, respectively. The stress
$\tau_{\rm r\varphi}=\alpha\sqrt{P_{\rm gas}P_{\rm tot}}$ is adopted.}\label{fig2}
\end{figure}

In Figure \ref{fig2}, we plot the interior temperatures of the
accretion disks as functions of radius $R$ with different values of
the accretion rate $\dot{m}$ and black hole spin parameter $a$, and the stress
$\tau_{\rm r\varphi}=\alpha\sqrt{P_{\rm gas}P_{\rm tot}}$ is adopted in the calculations. It
is not surprising that the temperature of the disk increases with
the mass accretion rate $\dot{m}$. For a rapidly spinning black
hole, the inner edge of the disk extends to the region close to the
black hole, and the temperature of the gas in the inner edge of the
disk is significantly higher than that for a slowly spinning (or
non-spinning) black hole.

%

The structures of the corona calculated with $\tau_{\rm
r\varphi}=\alpha\sqrt{P_{\rm gas}P_{\rm tot}}$, i.e., the
temperatures of the ions and electrons, and the electron scattering
optically depth of the corona in the vertical direction, are plotted
in Figure \ref{fig3}.
The electron temperature decreases with increasing mass accretion
rate $\dot{m}$ and the black hole spin parameter $a$. The optical
depth for electron scattering is in the range of $\sim0.1-0.3$ for
different values of $\dot{m}$ and $a$ adopted.

The emergent spectra of the accretion disk-corona systems
with $\tau_{\rm r\varphi}=\alpha\sqrt{P_{\rm gas}P_{\rm
tot}}$ observed at infinity with different disk model parameters
and viewing angles are plotted in Figure \ref{fig4}. In each panel,
one of these parameters is fixed at a certain value and the other
ones are taken as free parameters in the calculations of the
emergent spectra. The corona is optically thin and geometrically
thick, and therefore its radiation is almost isotropic. The
radiation of the disk is dominantly in the optical/UV bands, the
emergent spectrum of which depends on the viewing angle sensitively
due to its slab-like geometry.
The photon spectral indices, and the ratios of the bolometric
luminosity to the X-ray luminosity in $2-10$~keV as functions of the
accretion rate $\dot{m}$ for different values of $a$ and
$\theta_{\rm obs}$ are plotted in Figure \ref{fig5}. The hard X-ray
spectral index $\Gamma$ increases with the accretion rate $\dot{m}$
and black hole spin $a$, while it is insensitive with inclination
$\theta_{\rm obs}$. The ratio of the bolometric luminosity to the
X-ray luminosity increases with mass accretion rate, which is higher
for a rapidly spinning black hole.

The calculated emergent spectrum is usually insensitive to the outer
radius of the accretion disk, except the spectral shape in the
infrared/optical wavebands. The outer radius of the accretion disk
can be estimated by assuming the disk to be truncated where the
self-gravity of the disk dominates over the central gravity of the
black hole. The outer radii of the accretion disks with different
accretion rates $\dot{m}$ and black hole masses $M_{\rm bh}$ are
plotted in Figure \ref{fig6}. The results show that the accretion
disk surround a less massive black hole extends to a larger radius.
The spectral shape of the accretion disk in the infrared/optical
wavebands also varies with the outer radius of the disk $R_{\rm
out}$. We plot the emergent spectra of the disk-corona systems
with $\tau_{\rm r\varphi}=\alpha\sqrt{P_{\rm gas}P_{\rm
tot}}$ in the near-infrared/optical wavebands with different values
of the accretion rate $\dot{m}$ and black hole spin parameter $a$ in
Figure \ref{fig7}. We find that the spectral shapes of the accretion
disk-corona systems deviate significantly from
$f_{\nu}\varpropto\nu^{1/3}$, which is predicted by the standard
accretion disk model
\citep{1973A&A....24..337S,1999PASP..111....1K,2000ApJ...533..710H}.
\cite{2007ApJ...668..682D} measured UV spectral slopes ($s$, where
$f_{\nu}\varpropto\nu^{s}$) for several thousand quasars from the
Sloan Digital Sky Survey (SDSS), and the distributions of the slope
$s$ span a wide range rather than $s=1/3$ predicted by the standard
accretion model.

\section{Discussion}

In this work, we calculate the structure of an accretion disk with
corona around a Kerr black hole, and the emergent spectrum of the
system is derived with the ray tracing approach, in which the
general relativistic effects have been considered. The ions and
electrons in the corona are assumed to be heated by magnetic power
with the reconnection of the buoyant magnetic fields transported
from the disk \citep{1998MNRAS.299L..15D}. The strength of the
magnetic fields can be estimated from gas/radiation pressure of the disk with
assumed magnetic stress tensor. We examine the relations of the
ratio $L_{\rm cor}/L_{\rm bol}$ with the mass accretion rate
$\dot{m}$ in the models with different magnetic stress tensors (see
Figure \ref{fig1}). In order to understand the results, we analyze
the situation based on the standard thin disk model. The
gravitational power dissipated in unit surface area of the accretion
disk is given by \citep{1973A&A....24..337S}
\begin{equation}\label{dissi_viscosity}
Q_{\rm dissi}^{+}=\frac{1}{2}W_{\rm r\varphi}R\frac{d\Omega_{\rm
k}}{dr}=\frac{3}{2}\tau_{\rm r\varphi}H_{\rm d}\Omega_{\rm
k}=\frac{3}{2}\tau_{\rm r\varphi}c_{\rm s}=\frac{3}{2}\tau_{\rm
r\varphi}\left(\frac{p_{\rm tot}+p_{\rm m}}{\rm \rho}\right)^{1/2}
\end{equation}
where $W_{\rm r\varphi}=\int\limits_{-H_{\rm d}}^{H_{\rm
d}}\tau_{\rm r\varphi}dh$, $p_{\rm tot}=p_{\rm gas}+p_{\rm rad}$. The power dissipated in the corona is
\begin{equation}\label{cor_viscosity}
Q_{\rm cor}^{+}=p_{\rm m}\upsilon_{\rm p}=b\upsilon_{\rm A}p_{\rm
m}=\frac{bp_{\rm m}B}{({4\pi\rho})^{1/2}}=bp_{\rm m}\left({\frac
{2p_{\rm m}}{\rho}} \right)^{1/2}
\end{equation}
Combining Equations (\ref{dissi_viscosity}) and
(\ref{cor_viscosity}), we find that the ratio $L_{\rm cor}/L_{\rm
bol}\propto p_{\rm m}/(p_{\rm m}+p_{\rm tot})$. If the stress tensor
$\tau_{\rm r\varphi}=p_{\rm m}=\alpha p_{\rm tot}$ is adopted, the
ratio $L_{\rm cor}/L_{\rm bol}$ remains constant independent of
accretion rate $\dot{m}$. For the cases of $\tau_{\rm
r\varphi}=p_{\rm m}=\alpha p_{\rm gas}$, or $\tau_{\rm
r\varphi}=\alpha \sqrt{p_{\rm gas}p_{\rm tot}}$, the ratio $L_{\rm
cor}/L_{\rm bol}\propto T_{\rm d}^{-3}$ or $T_{\rm d}^{-3/2}$
respectively, as $p_{\rm gas}\sim T_{\rm d}$, $p_{\rm rad}\sim
T_{\rm d}^{4}$ and the radiation pressure is dominant in the inner
regions of the accretion disk. As the disk temperature increases
with mass accretion rate $\dot{m}$, the ratio $L_{\rm cor}/L_{\rm
bol}$ therefore decreases with increasing $\dot{m}$. The results
obtained in this work for spinning black holes are qualitatively
consistent with those derived in \citet{2009MNRAS.394..207C}.

\begin{figure}
\epsscale{1.25} \plotone{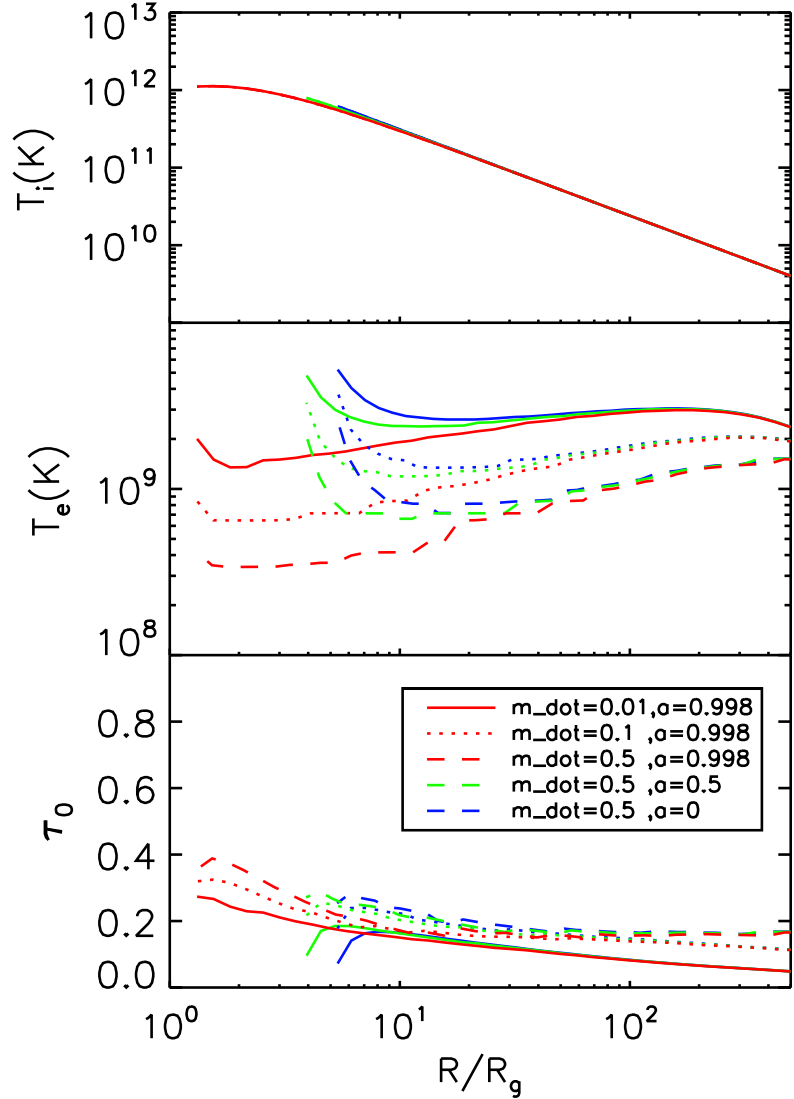} \caption{The upper panel: the
temperatures of the ions in the coronae as functions of radius $R$ with $\tau_{\rm r\varphi}=\alpha\sqrt{P_{\rm gas}P_{\rm tot}}$.
The middle panel: the temperatures of the hot electrons in the
coronae. The lower panel: the optical depth for the Compton
scattering of the electrons in the vertical direction of the
coronae. The colored lines represent different values of black hole
spin parameter, i.e., $a=0.998$(red), 0.5 (green), and 0 (blue). The
solid lines represent the results calculated with $\dot{m}=0.01$,
while the dotted and dash lines are for $\dot{m}=0.1$ and $0.5$,
respectively. }\label{fig3}
\end{figure}

\begin{figure}
\epsscale{1.25} \plotone{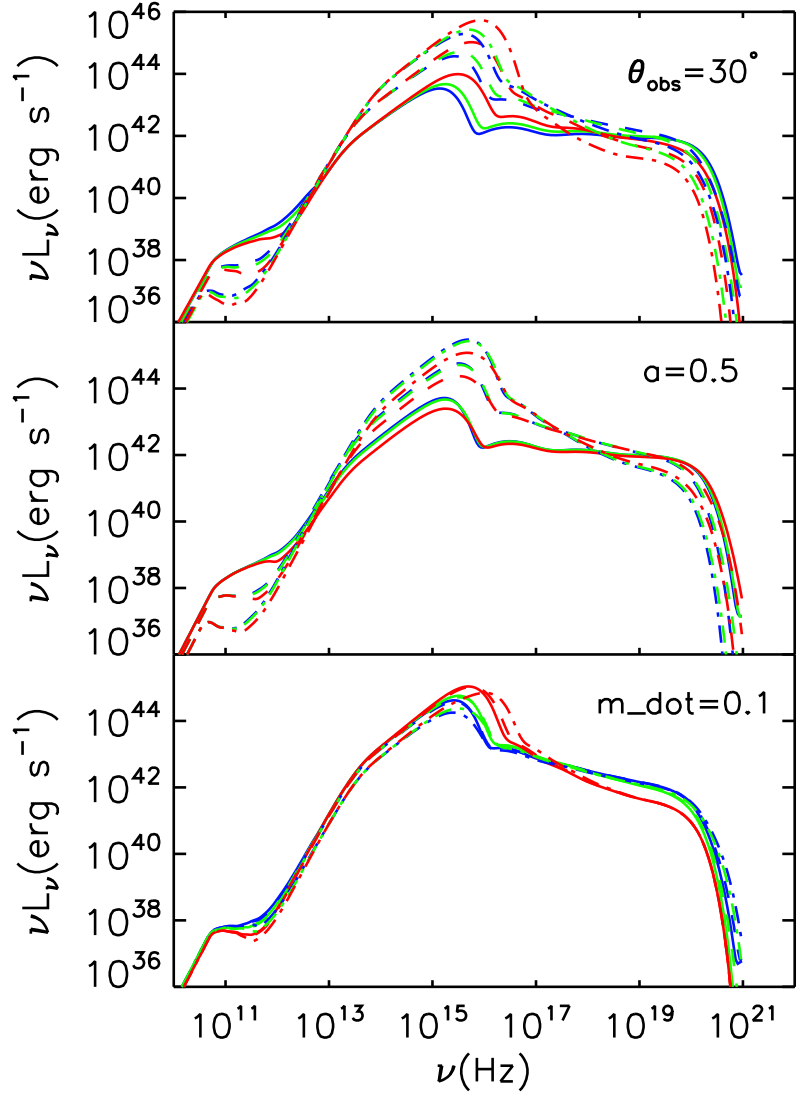} \caption{The spectra of the
disk-corona systems with different values of $\dot{m}$, $a$, and
$\theta_{\rm obs}$. The upper panel: the emergent spectra of the
disk-corona systems surrounding spinning black holes observed at an
angle $\theta_{\rm obs}=30^{\circ}$ with respect to the disk axis
(red: $a=0.998$, green: $a=0.5$, and blue: $a=0$). The different
type lines represent the spectra calculated with different values of
accretion rate, i.e., $\dot{m}=0.5$ (dash-dotted), $0.05$ (dashed),
and $0.01$ (solid). The middle panel: the spectra of the disk-corona
systems surrounding black holes spinning at $a=0.5$. The different
colored lines represent the emergent spectra observed at different
inclination angles, i.e., $\theta_{\rm obs}=60^{\circ}$ (red),
$30^{\circ}$ (green), and $10^{\circ}$ (blue). The different type
lines represent the different values of accretion rate, i.e.,
$\dot{m}=0.5$ (dash-dotted), $0.05$ (dashed), and $0.01$ (solid).
The bottom panel: the spectra of the disk-corona systems calculated
with a given accretion rate, $\dot{m}=0.1$. The different colored
lines represent the spectra with different values of black hole spin
parameter, i.e., $a=0.998$ (red), $0.5$ (green), and $0$ (blue). The
spectra observed at different angles are plotted with different type
lines, i.e., $\theta_{\rm obs}=60^{\circ}$ (dash-dotted),
$30^{\circ}$ (dashed), and $10^{\circ}$ (solid). The stress
$\tau_{\rm r\varphi}=\alpha\sqrt{P_{\rm gas}P_{\rm tot}}$ is adopted in the calculations.}\label{fig4}
\end{figure}

\begin{figure}
\epsscale{1.25} \plotone{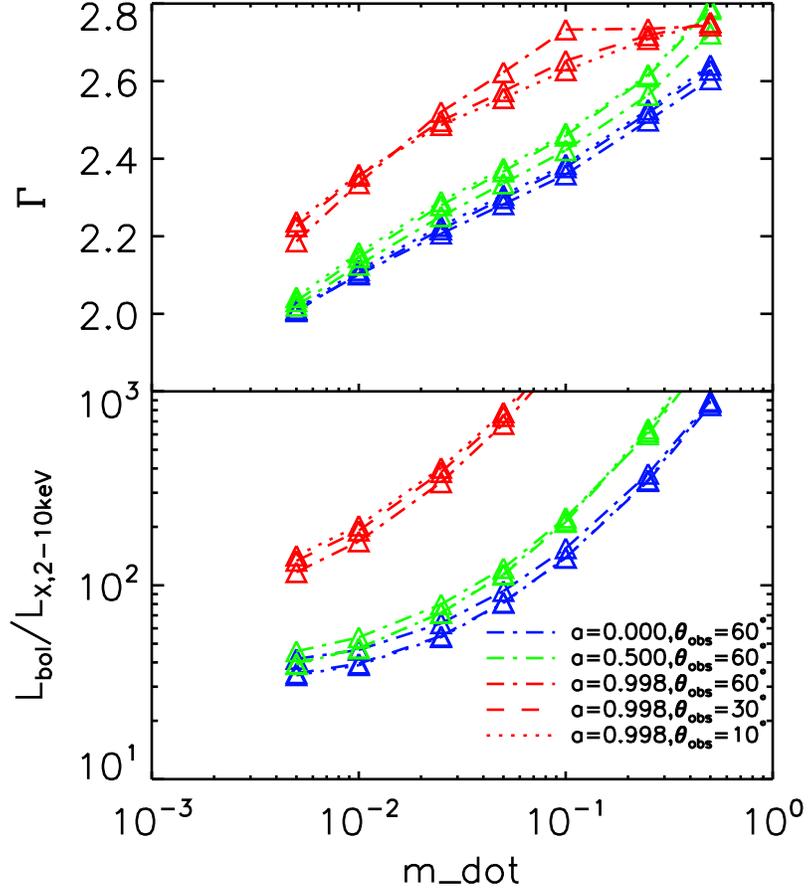} \caption{The upper panel: the
photon spectral indices as functions of accretion rate $\dot{m}$ for
different values of black spin parameter $a$ and viewing angle
$\theta_{\rm obs}$ with $\tau_{\rm r\varphi}=\alpha\sqrt{P_{\rm gas}P_{\rm tot}}$. The lower panel: the X-ray bolometric correction
factors $L_{\rm bol}/L_{\rm X,2-10keV}$ as functions of accretion
rate $\dot{m}$ for different model parameters. The colored lines
represent the results calculated with $a=0.9$ (red), 0.5 (green),
and 0 (blue), respectively. The different type lines represent the
spectra observed at different angles, i.e., $\theta_{\rm
obs}=60^{\circ}$ (dash-dotted), $30^{\circ}$ (dashed), $10^{\circ}$
(dotted). } \label{fig5}
\end{figure}

\begin{figure}
\epsscale{0.9} \plotone{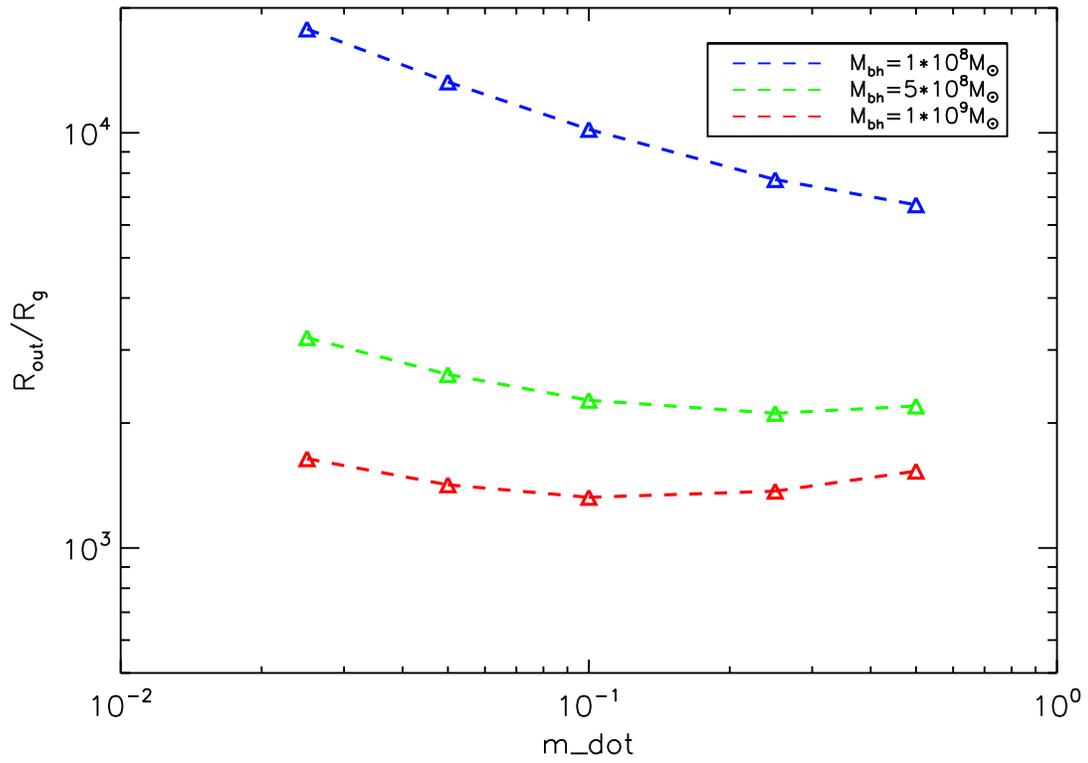} \caption{The outer radii of the
accretion disks as functions of accretion rate $\dot{m}$ for
different black hole mass $M_{\rm bh}$ with $\tau_{\rm r\varphi}=\alpha\sqrt{P_{\rm gas}P_{\rm tot}}$(blue: $M_{\rm bh}=10^8
M_{\odot}$, green: $M_{\rm bh}=5\times 10^8 M_{\odot}$, red: $M_{\rm
bh}=10^9 M_\odot$).  } \label{fig6}
\end{figure}

\begin{figure}
\epsscale{0.9} \plotone{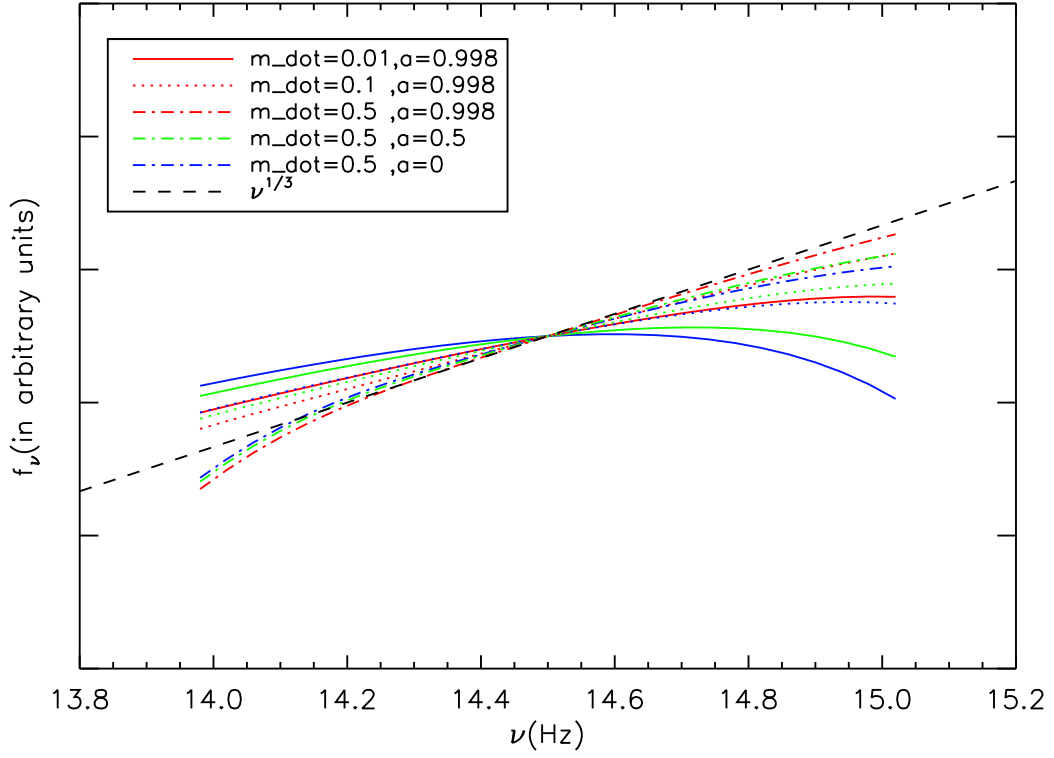} \caption{The spectral shapes of
the disk-corona systems in the near IR/optical wavebands with $\tau_{\rm r\varphi}=\alpha\sqrt{P_{\rm gas}P_{\rm tot}}$. The
colored lines represent the model calculations with different values
of black hole spin parameter, i.e., $a=0$ (blue), $0.5$ (green), and
$0.998$ (red). The different type lines represent the results
calculated with different values of accretion rate, i.e.,
$\dot{m}=0.01$ (solid), $0.1$ (dotted), and $0.5$ (dash-dotted). The
dashed line represents $f_{\nu}\varpropto\nu^{1/3}$, expected by the
standard thin disk model. } \label{fig7}
\end{figure}

As the value of black hole spin parameter $a$ increases, the disk
extends close to the black hole, and more gravitational power is
released in the inner region of the accretion disk. The temperature
of the inner disk region $T_{\rm d}$ increases with the spin
parameter $a$, which increases the energy density of the soft
photons emitted from the disk. This leads to the corona cooling
efficiently, because the cooling of the corona is dominated by the
Compton scattering of the soft photons from the disk. The electron
temperature of the corona in the inner region of the disk is
therefore lower for rapidly spinning black holes (see Figure
\ref{fig3}).

The resulted spectra of the disk-corona systems with magnetic stress
tensor $\tau_{r\varphi}=\alpha\sqrt{p_{\rm gas}p_{\rm tot}}$ show
that both the hard X-ray spectral index and the hard X-ray
bolometric correction factor $L_{\rm bol}/L_{\rm X,2-10 keV}$
increase with the Eddington ratio, which are qualitatively
consistent with the observations of AGNs
\citep*[][]{2004ApJ...607L.107W,2006ApJ...646L..29S,2007MNRAS.381.1235V},
and X-ray binaries \citep{2008ApJ...682..212W}. The observed
ratio of $L_{\rm opt}/L_{\rm bol}$ may provide additional
information of the disk-corona systems \citep{2011ApJ...728...98D}.
The stress tensor $\tau_{r\varphi}=\alpha \sqrt{p_{\rm gas}p_{\rm
tot}}$ was initially suggested by \citet{1984ApJ...287..761T} based
on the idea that the viscosity is proportional to the gas pressure,
but the size of turbulence should be limited by the disk scale
height determined by the total pressure. This is also supported by
the analysis on the local dynamical instabilities in magnetized,
radiation-pressure-supported accretion disks
\citep{2001ApJ...553..987B,2002MNRAS.332..165M}. However, recent
radiative MHD simulations seem to prefer the stress tensor
$\tau_{r\varphi}=\alpha p_{\rm tot}$ \citep{2009ApJ...704..781H}.
Our model calculations can be carried out when the form of magnetic
stress tenor is specified. For simplicity, in most cases of this
work, we calculate the structure and emergent spectra of accretion
disk-corona systems using the magnetic stress tensor
$\tau_{r\varphi}=\alpha\sqrt{p_{\rm gas}p_{\rm tot}}$.



The radiation from the corona is almost isotropic due to its
optically thin and geometrically thick structure, while the
radiation from the disk is anisotropic. It is therefore found that
the hard X-ray spectral index $\Gamma$ is insensitive with the
viewing angle $\theta_{\rm obs}$. The X-ray spectra are mainly
dominated by the inverse Compton scattering, which mainly depends on
the temperature of the electrons. Therefore the trend of the photon
spectral indices $\Gamma$ and the X-ray correction factors $L_{\rm
bol}/L_{\rm X,2-10keV}$ can be explained by the decrease of the
temperature of the electrons with increasing mass accretion rate
$\dot{m}$, or the black hole spin parameter $a$ (see Figure
\ref{fig5}, and also Figure \ref{fig3}). The photon index $\Gamma$
of the hard X-ray spectrum becomes large when the black
hole is rapidly spinning if all other parameters are fixed, which is
due to the temperature of the electrons decreases with increasing
value of $a$. The correction factor $L_{\rm bol}/L_{\rm
X,2-10keV}$ derived by the model calculations with
$\tau_{r\varphi}=\alpha\sqrt{p_{\rm gas}p_{\rm tot}}$ are
significantly larger than the observed values when $\dot{m}\sim0.1$
\citep*[e.g.,][]{2004ApJ...607L.107W,2007MNRAS.381.1235V}. One
possibility is that an advection dominated accretion flow (ADAF) is
present in the inner region of the disk
\citep{1994ApJ...428L..13N,1995ApJ...452..710N}. The inner ADAF is
X-ray luminous, which leads to a low value of $L_{\rm bol}/L_{\rm
X,2-10keV}$ \citep*[][]{1999ApJ...525L..89Q}. This is also
consistent with the fact that the X-ray spectra of the AGNs are hard
when accreting at low rates \citep*[see][for the detailed
discussion]{2004ApJ...607L.107W,2009MNRAS.394..207C}. An alternative
resolution is the model with magnetic stress tenor
$\tau_{r\varphi}=\alpha p_{\rm tot}$, however, it always predicts a
constant $L_{\rm bol}/L_{\rm X,2-10keV}$ independent of mass
accretion rate $\dot{m}$. This needs further investigations in the
future.


In Figure \ref{fig7}, the spectral shapes of the disk-corona model
with different values of accretion rates $\dot{m}$ and black hole
spin parameter $a$ in the near-infrared/optical bands are plotted,
which is dominantly contributed by the blackbody radiation from the
disk. The spectral shapes are significantly different from
$f_{\nu}\varpropto\nu^{1/3}$, as expected in standard thin disk
model. This is mainly caused by the $R$-dependent temperature
distributions in the disks surrounding massive black holes and the
outer radii of the disks being different from those in standard thin
disk model. In the standard thin disk model, the spectral shape
$f_{\nu}\varpropto\nu^{1/3}$ in the near-infrared/optical wavebands
is derived by assuming the temperature of the accretion disk $T_{\rm
d}\sim R^{-3/4}$ and the disk to extend to infinity. However, the
temperature profiles of relativistic accretion disks deviate
significantly from $T_{\rm d}\sim R^{-3/4}$ in the inner region of
the disk ($R<100R_{\rm S}$) (see Figure \ref{fig2}). The
contribution to the spectra in IR/optical bands from the blackbody
radiation of the outer region of the disks can also be quite
important. The outer radius of the disk is a function of black hole
mass $M_{\rm bh}$ and accretion rate $\dot{m}$ (see Figure
\ref{fig6}).
The observational slope of the infrared continua can be used to
constrain the accretion disk model
\citep*[see][]{2008Natur.454..492K}. It is not suitable to compare
the observed spectra only with, $f_{\nu}\varpropto\nu^{1/3}$, as
predicted by the standard accretion disk model.
In this work, the local radiation from the disk is
approximated as blackbody emission. A better approximation is to
include a spectral hardening factor in calculating the local
spectrum of the disk
\citep*[e.g.,][]{1995ApJ...445..780S,2000ApJ...533..710H,2007ApJ...668..682D},
with which the energy peak of the calculated disk spectrum will
shift to a slightly higher frequency, while the spectral shape
changes little. This would be important for detailed fitting the
observed SED of AGNs, which is beyond the scope of this paper.

It is assumed that zero torque at the inner edge of the disk
\citep{1973A&A....24..337S}, which was doubted in some previous
works \citep{2000ApJ...528..161A,2001ApJ...548..348H}. The
general relativistic magnetohydrodynamic (GRMHD) simulations provide
an effective tool to explore the torque in the plunging region
\citep{2008ApJ...687L..25S,2011MNRAS.414.1183K}. In the accretion
disk with the thickness $h/r\sim0.05-0.1$ the magnetic torque at the
radius of the innermost stable circular orbit (ISCO) is only $\sim
2\%$ of the inward flux of angular momentum at this radius, which
indicates that the zero torque is really a good approximation for
geometrically thin disks \citep{2008ApJ...687L..25S}.
\cite{2012MNRAS.420.1415G} investigated the accretion disk-corona
model similar to that developed by \citet{2009MNRAS.394..207C}, in
which a non-zero torque at the inner edge of the disk is assumed. As
the detailed accretion physics in the plunging region is complicated
and still quite unclear, we adopt the assumption of no torque in the
inner edge of the accretion disk in all our calculations. The
calculations in this work were carried out for massive black holes,
which will be compared with the observed spectra of AGNs in our
future work. The model calculations can also be applicable to X-ray
binaries.

\section*{Acknowledgements}
This work is supported by the National Basic Research Program of
China (grant 2009CB824800, 2012CB821800), the NSFC (grants grants
11173043, 11121062, 11233006, 11073020, and 11133005), the CAS/SAFEA
International Partnership Program for Creative Research Teams
(KJCX2-YW-T23), and the Fundamental Research Funds for the Central
Universities (WK2030220004).

.

{}

\clearpage

\clearpage

\clearpage

\clearpage

\clearpage

\end{document}